\RequirePackage{fix-cm}
\documentclass[smallextended]{svjour3}       
\usepackage{graphicx}
\usepackage[misc]{ifsym}
\DeclareUnicodeCharacter{2212}{-}
\begin{document}

\title{Stability and Horizon Formation during Dissipative Collapse}

\author{Nolene F. Naidu$^\dag$,
Robert S. Bogadi$^\ddag$, Anand Kaisavelu$^
\ast$, Megan Govender$^{\ast\ast}$}%


\institute{\at
			  Department of Mathematics, Faculty of Applied Sciences, Durban University of Technology, Durban, 4000, South Africa. \\
              \email{nolene.naidu@physics.org$^\dag$ $^{\textrm{\Letter}}$}             \\
              \email{bogadi.robert@gmail.com$^\ddag$}\\
              \email{megandhreng@dut.ac.za$^{\ast\ast}$}   \\
			  Astrophysics and Cosmology Research Unit, School of Mathematics, Statistics and Computer Science, University of KwaZulu-Natal, Private
			  Bag X54001, Durban 4000, South Africa. \\
			  \email{anand.kaisavelu@gmail.com$^\ast$}
 }

\date{Received: date / Accepted: date}
\maketitle
\begin{abstract}
We investigate the role played by density inhomogeneities and dissipation on the final outcome of collapse of a self-gravitating sphere. By imposing a perturbative scheme on the thermodynamical variables and gravitational potentials we track the evolution of the collapse process starting off with an initially static perfect fluid sphere which is shear-free. The collapsing core dissipates energy in the form of a radial heat flux with the exterior spacetime being filled with a superposition of null energy and an anisotropic string distribution. The ensuing dynamical process slowly evolves into a shear-like regime with contributions from the heat flux and density fluctuations. We show that the anisotropy due to the presence of the strings drives the stellar fluid towards instability with this effect being enhanced by the density inhomogeneity. An interesting and novel consequence of this collapse scenario is the delay in the formation of the horizon.
\end{abstract}


\section{Introduction}
Gravitational collapse is fundamental to the formation of the majority of stellar objects in the universe and thus one would expect that the study
of this phenomenon should be vital to our understanding of the workings of the universe. The pioneers in the research of gravitational collapse,
Oppenheimer and Snyder (1939), studied a spherically symmetric matter distribution in the form of a dust sphere undergoing collapse. They obtained
the first solution for the non-adiabatic collapse of a dust ball with a Schwarzschild exterior \cite{oppsnyd}. Vaidya \cite{vaidya} obtained an
exact solution to the Einstein field equations which describes the exterior field of a radiating, spherically symmetric fluid by noting that a
radiating collapsing mass distribution has outgoing energy and so its exterior spacetime is no longer a vacuum but contains null radiation. The
next step in improving the model was accomplished by Santos \cite{santos} who derived the junction conditions for a collapsing spherically
symmetric, shear-free non-adiabatic fluid sphere with heat flow. The combination of these contributions allowed for the matching of the interior
and exterior spacetimes of a collapsing star which lead the way for studying non-adiabatic, isotropic as well as anisotropic dissipative
gravitational collapse \cite{mken}-\cite{kevin}.

A disturbance or perturbation of a system initially in static equilibrium results in a change in stability which likely renders the system dynamic.
The property of a system to retain its initial stable state once perturbed, is then referred to as its dynamical (in)stability. Hence the issue of
stability is vital in the study of self-gravitating objects as a static stellar model which evolves towards higher instability is of little physical
significance. The dynamical instability of a spherically symmetric mass with isotropic pressure was first investigated by
Chandrasekhar \cite{chandra}. He showed that for a system to remain stable under collapse, the adiabatic index $\Gamma$ must be greater
than $\frac{4}{3}$.

Subsequently, Herrera et al. \cite{herr2} showed that for a non-adiabatic sphere where relativistic corrections were imposed to address heat flow,
the unstable range of $\Gamma$ decreased rendering the fluid less unstable. Chan et al.  \cite{chan2} investigated the stability criteria by
deviating from the perfect fluid condition in two ways: they considered radiation in the free-streaming approximation; and they assumed local
anisotropy of the fluid. Herrera et al. \cite{herr3} also examined the dynamical instability of expansion-free, locally anisotropic spherical
stellar bodies.

The application of Einstein's field equations to increasingly more complex gravitating systems with additional parameters and degrees of freedom
depends on computational techniques, which is the case when perturbative theories are employed in which higher order terms arise. The generalization
of systems, such as the inclusion of a string density field, can increase the complexity of expressions obtained however to first order we aim to
introduce the temporal behaviour and hence evolution of the collapse process. This method is well established \cite{chan2,bon}. Compact objects
such as neutron stars, black holes and the more recently proposed dark-fluid stars and strange stars composed of quark matter invite the addition
of a more complex, non-empty stellar exterior. The Vaidya metric which is commonly used for describing the exterior spacetime would then require
modification to include both the radiation field and a so-called string field as initially put forward by Glass and Krisch \cite{glass,glass2}.
In this more generalized Vaidya exterior, the mass function is augmented to acquire both temporal and spatial dependence.

In 2005, Maharaj and Govender showed that the stellar core was more unstable than the outer regions by investigating gravitational collapse with
isotropic pressure and vanishing Weyl stresses \cite{maharaj2}. More recently, Maharaj et al. \cite{maharaj3} showed the impact of the generalized
Vaidya radiating metric on the junction conditions for the boundary of a radiating star. Their results describe a more general atmosphere
surrounding the star, described by the superposition of a pressure-free null dust and a string fluid. The string density was shown to affect
the fluid pressure at the surface of the star. It was demonstrated that this string density reduces the pressure at the stellar boundary. The
usual junction conditions for the Vaidya spacetime are regained in the absence of the string fluid.

In this study, a spherically symmetric static configuration undergoing radiative collapse under shear-free, isotropic conditions is considered.
A boundary condition of the form $\left(p_r\right)_{\Sigma} = \left(qB\right)_{\Sigma} - \rho_s$ is imposed where $\rho_s$ is the string density
and $qB$ is the heat flux.
This is the basis for developing the temporal behaviour of the self-gravitating system.
The structure of this paper is as follows: In $\S$2 the field equations describing the geometry and matter content for a star undergoing
shear-free gravitational collapse are introduced. In $\S$3 the exterior spacetime and the junction conditions necessary for the smooth matching
of the interior spacetime with Vaidya's exterior solution across the boundary are presented. In $\S$4 the perturbative scheme is described and the
field equations for the static and perturbed configurations are stated. In $\S$5 we develop the new temporal equation employed in the perturbative
scheme which includes the effect of the string field. In $\S$6, we develop a radiating model from an interior Schwarzschild static model. In $\S$7
dissipative collapse is discussed, the perturbed quantities in terms of two unspecified quantities are expressed and an equation of state which
presents the perturbed quantities in terms of radial coordinate only is introduced. The stability of the collapsing model in the Newtonian and
post-Newtonian approximations are explored in $\S$8. The physical analysis of the results and conclusion are discussed in $\S$9. The
acknowledgements follows in $\S$10.

\section{Stellar Interior}
\label{interior}
In order to investigate the evolution of the radiative collapse we adopt a spherically symmetric shear-free line element in simultaneously
comoving and isotropic coordinates given by
\begin{equation}
ds_-^2 = -A^2(r,t)dt^2 + B^2(r,t)[dr^2 + r^2(d\theta^2 + \sin^2{\theta}d\phi^2)]  \label{metric}
\end{equation}
where the $A(r,t)$ and $B(r,t)$ are the dynamic gravitational potentials. We should highlight the fact that the stability of the shear-free
condition may hold for a limited period of the collapse process \cite{chan2}. Herrera and co-workers have shown that shear-free collapse can evolve
into a dynamical process mimicking shear. The shear-like contributions can develop from pressure anisotropy and density inhomogeneities.
The stellar material for the interior is described by an imperfect fluid with heat flux and in the form of the energy-momentum tensor
\begin{equation}
 T_{ab} = (\rho + p_t)u_a u_b + p_tg_{ab} + (p_r - p_t) \chi_a \chi_b + q_a u_b + q_b u_a \label{2}
\end{equation}
where $\rho$ is the energy density, $p_r$ the radial pressure, $p_t$ the tangential pressure and $q_a$ the heat flux vector, $u_a$ is the
timelike four-velocity of the fluid and $\chi_a$ is a spacelike unit four-vector along the radial direction. These quantities must satisfy
$u_a u^a = -1$, $u_a q^a = 0$, $\chi_a \chi^a = 1$ and $\chi_a u^a = 0$.
In co-moving coordinates we must have
\begin{eqnarray}
u^a&=&A^{-1}\delta^a_0 \label{k2a}
\end{eqnarray}
\begin{eqnarray}
q^a&=&q\delta^a_1 \label{k2b}
\end{eqnarray}
\begin{eqnarray}
\chi^a&=&B^{-1}\delta^a_1 \label{k2c}
\end{eqnarray}
The nonzero components of the Einstein field equations for line element (\ref{metric}) with energy-momentum tensor (\ref{2}) are
\begin{eqnarray} \label{3}
\rho &=&  - \frac{1}{B^2}\left[2\frac{B^{\prime\prime}}{B} - \left(\frac{B^{\prime}}{B}\right)^2 +\frac{4}{r}\frac{B^{\prime}}{B}\right] +\frac{3}{A^2}\left(\frac{{\dot B}}{B}\right)^2 \label{3a}  \\
p_r &=& \frac{1}{B^2}\left[\left(\frac{B^{\prime}}{B}\right)^2 + \frac{2}{r}\left(\frac{A^{\prime}}{A}+ \frac{B^{\prime}}{B}\right) + 2\frac{A^{\prime}}{A}\frac{B^{\prime}}{B}\right] \nonumber \\ \nonumber \\
&&+ \frac{1}{A^2}\left[-2\frac{\ddot B}{B} - \left(\frac{\dot B}{B}\right)^2 + 2\frac{\dot A}{A}\frac{\dot B}{B}\right] \label{3b} \\
p_t &=& \frac{1}{A^2}\left[-2\frac{\ddot B}{B} -\left( \frac{\dot B}{B}\right)^2 +2 \frac{\dot A}{A}\frac{\dot B}{B}\right] \nonumber \\ \nonumber \\
&& + \frac{1}{B^2}\left[\frac{B^{\prime\prime}}{B}-\left(\frac{B^{\prime}}{B}\right)^2+\frac{1}{r}\left(\frac{A^{\prime}}{A}+\frac{B^{\prime}}{B}\right)+\frac{A^{\prime\prime}}{A}\right]
\label{3c} \\
q &=& \frac{2}{AB^2}\left[\frac{{\dot B}^{\prime}}{B}-\frac{\dot B}{B}\left(\frac{B^{\prime}}{B}+\frac{A^{\prime}}{A}\right)\right]
\label{3d}
\end{eqnarray}
where dots and primes represent partial derivatives with respect to $t$ and $r$ respectively.

\section{Exterior Spacetime and Matching Conditions}
\label{exterior}
Since the star is radiating, the exterior spacetime can be described by the generalized Vaidya metric \cite{vaidya} which represents a mixture of null radiation and strings \cite{glass}
\begin{equation}\label{gvmetric}
ds^2_+ = -\left[1 - \frac{2m(v, \mathrm{r})}{\mathrm{r}}\right]dv^2 - 2dvd\mathrm{r} + \mathrm{r}^2(d\theta^2 + \sin^2{\theta}d\phi^2)\end{equation}
where $m(v, \mathrm{r})$ is the mass function which represents the total energy within a sphere of radius $\mathrm{r}$.
This is what distinguishes the generalized Vaidya solution from the pure radiation Vaidya solution, which has $m = m(v)$ where $v$ is
the retarded time.
The energy momentum tensor corresponding to line element (\ref{gvmetric}) is
\begin{equation}
T^+_{ab}=\tilde\mu l_al_b+(\rho+P)(l_an_b+l_bn_a)+Pg_{ab} \label{EnMom}
\end{equation}
where
\begin{equation}
l_a=\delta^{0}_{a}
\end{equation}
\begin{equation}
n_a=\frac{1}{2}\left[1-2\frac{m(v,\mathrm{r})}{\mathrm{r}}\right]\delta^{0}_{a}+\delta^{1}_{a}
\end{equation}
are null vectors such that $l_a l^a = n_a n^a = 0$ and $l_a n^a = −1$. The energy momentum tensor (\ref{EnMom}) can be interpreted as the matter source for the exterior atmosphere of the star which is a superposition of pressureless null dust and anisotropic null strings \cite{HusV,WangA}. The energy density of the null dust radiation, string energy density and string pressure are characterised by $\tilde \mu$, $\rho$ and $P$ respectively. We assume that the string  diffusion is equivalent to point particle diffusion where the number density diffuses from higher to lower numbers subjected to the continuity equation
\begin{equation} \label{diff}
\dot{\rho} = \frac{\cal{D}}{\mathrm{r}^2} \frac{\partial}{\partial \mathrm{r}} \left(\mathrm{r}^2 \frac{\partial \rho}{\partial \mathrm{r}}\right)
\end{equation}
where $\cal D$ is the positive coefficient of self-diffusion \cite{govin}.
Following de Oliveira et al. \cite{oliv2}, we obtain the boundary conditions which include a string density $\rho_s$,
\begin{equation}
\hspace{1cm} \left(p_r\right)_{\Sigma} = \left(qB\right)_{\Sigma} - \left(\rho_s\right)_\Sigma \label{4a}
\end{equation}
\begin{equation}
\hspace{1cm} \left(qB\right)_{\Sigma} = -\left(\frac{2}{{\mathrm{r}}^2}{{\dot v}^2}\frac{dm}{dv}\right)_{\Sigma} \label{4b}
\end{equation}
\begin{equation}
\hspace{1cm} \left(rB\right)_{\Sigma} = \mathrm{r}_{\Sigma} \label{4c}
\end{equation}
\begin{equation}
\hspace{1cm} \left(A dt\right)_{\Sigma} = \left(1 - \frac{2m}{\mathrm{r}} + 2\frac{d\mathrm{r}}{dv}\right)^{1/2}_{\Sigma} dv \label{4e}
\end{equation}
Equation (\ref{4a}) represents the conservation of momentum flux across the stellar boundary which we will employ in $\S$5 to determine the temporal evolution of our model. The total energy entrapped within a radius $r$ inside $\Sigma$ is given by
\begin{equation} \label{masss1}
m(r,t) = \frac{r^3B{\dot B}^2}{2A^2} - r^2{B^{\prime}} - \frac{r^3{{B^{\prime}}^2}}{2B}
\end{equation}
At the boundary, this is given by
\begin{equation}
\hspace{1cm} m(v,\mathrm{r}) = m(r,t) |_\Sigma
\end{equation}
and included as a boundary condition.

\section{Perturbative Scheme}
\label{pert}
Following the method in Herrera et al. \cite{herr2}, as well as the works of Chan et. al. \cite{chan2} and Govender et. al. \cite{gov2}, we present
our model in this section. To begin, we will assume that the fluid is in static equilibrium. The system is then perturbed and undergoes slow
shear-free dissipative collapse. Thermodynamical quantities in the static system are represented by a zero subscript, while those in the perturbed
fluid are represented by an overhead bar. The metric functions $A(r,t)$ and $B(r,t)$ are taken to have the same temporal dependence, which extends
to the perturbed material quantities. The time-dependent metric functions and material quantities are given by
\begin{eqnarray} \label{5}
A(r,t) &=& A_0(r) + \epsilon a(r)T(t) \label{5a} \\
B(r,t) &=& B_0(r) + \epsilon b(r)T(t) \label{5b} \\
\rho(r,t) &=& \rho_0(r) + \epsilon \bar \rho(r,t) \label{5c} \\
p_r(r,t) &=& p_{r0}(r) + \epsilon {\bar p}_r(r,t) \label{5d} \\
p_t(r,t) &=& p_{t0}(r) + \epsilon {\bar p}_t(r,t) \label{5e} \\
m(r,t) &=& m_0(r) + \epsilon \bar m(r,t) \label{5f} \\
q(r,t) &=& \epsilon \bar q(r,t) \label{5g}
\end{eqnarray}
where we assume that $0 < \epsilon << 1$. We observe that the temporal dependence of the perturbative quantities, $T(t)$ is the same for both  the gravitational potentials and the thermodynamical variables. The imposition of spherical
symmetry alone implies that we have a very large gauge (coordinate) freedom to write
the line element. In adopting the form of the line element given by (\ref{metric}) we exhaust all
coordinate freedom with the exception of re-scaling the radial
coordinate and/or the temporal coordinates. It is clear that such re-scaling would not change the form of (\ref{5a})-(\ref{5g}). The choice
of the perturbed variables as given in the perturbative scheme is not unique. However, once the
line element has been chosen, the choice of the perturbed variables cannot be varied
to produce the same physics \cite{kevin1}. \\

The Einstein field equations for the static configuration are given by
\begin{eqnarray} \label{6}
\rho_0 &=&  - \frac{1}{B_0^2}\left[2\frac{B_0^{\prime\prime}}{B_0} - \left(\frac{B_0^{\prime}}{B_0}\right)^2 +\frac{4}{r}\frac{B_0^{\prime}}{B_0}\right] \label{6a} \\
p_{r0} &=&  \frac{1}{B_0^2}\left[\left(\frac{B_0^{\prime}}{B_0}\right)^2 + \frac{2}{r}\left(\frac{A_0^{\prime}}{A_0}+ \frac{B_0^{\prime}}{B_0}\right) + 2\frac{A_0^{\prime}}{A_0}\frac{B_0^{\prime}}{B_0}\right] \label{6b} \\
p_{t0} &=&  \frac{1}{B_0^2}\left[\frac{B_0^{\prime\prime}}{B_0}-\left(\frac{B_0^{\prime}}{B_0}\right)^2+\frac{1}{r}\left(\frac{A_0^{\prime}}{A_0}+\frac{B_0^{\prime}}{B_0}\right)+\frac{A_0^{\prime\prime}}{A_0}\right] \label{6c}
\end{eqnarray} \\
The perturbed field equations up to first order in $\epsilon$ can be written as
\begin{eqnarray}
\bar \rho &=& -3\rho_0\frac{b}{B_0}T + \frac{1}{B_0^3}\left[-\left(\frac{B_0^{\prime}}{B_0}\right)^2 b + 2\left(\frac{B_0^{\prime}}{B_0} - \frac{2}{r}\right)b^{\prime} - 2b^{\prime\prime}\right]T \label{7a} \\
\bar p_r &=& -2p_{r0}\frac{b}{B_0}T + \frac{2}{B_0^2}\bigg[\left(\frac{B_0^{\prime}}{B_0} + \frac{1}{r} + \frac{A_0^{\prime}}{A_0}\right)\left(\frac{b}{B_0}\right)^{\prime}  \nonumber \\
&& + \left(\frac{B_0^{\prime}}{B_0} + \frac{1}{r}\right)\left(\frac{a}{A_0}\right)^{\prime}\bigg]T - 2\frac{b}{{A_0^2}B_0}{\ddot T}  \label{7b}  \\
\bar p_t &=&  -2p_{t0}\frac{b}{B_0}T + \frac{1}{B_0^2}\bigg[\left(\frac{b}{B_0}\right)^{\prime\prime} + \frac{1}{r}\left(\frac{b}{B_0}\right)^{\prime} + 2\frac{A_0^{\prime}}{A_0}\left(\frac{a}{A_0}\right)^{\prime}   \nonumber \\
&& + \left(\frac{a}{A_0}\right)^{\prime\prime} + \frac{1}{r}\left(\frac{a}{A_0}\right)^{\prime}\bigg]T - 2\frac{b}{{A_0^2}B_0}{\ddot T}  \label{7c} \\
\bar q B &=& \frac{2}{B_0}\left(\frac{b}{A_0 B_0}\right)^{\prime}{\dot T} \label{7d}
\end{eqnarray}
The total energy enclosed within $\Sigma$ is obtained by using (\ref{masss1}) and (\ref{5f}). We separate the static and time-dependent/perturbed components
and are shown as follows
\begin{eqnarray} \label{8}
m_0(r_{\Sigma}) &=&  -\left( r^2{B_0^{\prime}} + \frac{r^3{{B_0^{\prime}}^2}}{2B_0}\right)_{\Sigma} \label{8a}
\end{eqnarray}
and
\begin{eqnarray}
\bar m(r_{\Sigma},t) &=& -\left(\left[{r^2}b^{\prime} + \frac{r^3{{B_0^{\prime}}^2}}{2B_0} \left(2\frac{b^{\prime}}{B_0^{\prime}}-\frac{b}{B_0}\right)\right]T(t)\right)_{\Sigma} \label{8b}
\end{eqnarray}
In the case where the radial and tangential stresses are equal, $p_r = p_t$, the condition of pressure isotropy for the static model is
$p_{r0} = p_{t0}$ which gives
\begin{eqnarray}
\left(\frac{A'_0}{A_0} + \frac{B'_0}{B_0}\right)^\prime - \left(\frac{A'_0}{A_0} + \frac{B'_0}{B_0}\right)^2  - \frac{1}{r}\left(\frac{A'_0}{A_0} + \frac{B'_0}{B_0}\right)
 + 2\left(\frac{A'_0}{A_0}\right)^2 = 0
\end{eqnarray}
The pressure isotropy condition for the perturbed model is $\bar p_r = \bar p_t$ which gives
\begin{eqnarray}
&& \left[\left(\frac{a}{A_0}\right)^\prime + \left(\frac{b}{B_0}\right)^\prime\right]^\prime - 2\left[\left(\frac{a}{A_0}\right)^\prime + \left(\frac{b}{B_0}\right)^\prime\right]\left(\frac{A'_0}{A_0} + \frac{B'_0}{B_0}\right) \nonumber \\
&& - \frac{1}{r}\left[\left(\frac{a}{A_0}\right)^\prime + \left(\frac{b}{B_0}\right)^\prime\right] + 4\frac{A'_0}{A_0}\left(\frac{a}{A_0}\right)^\prime = 0  \label{pertiso}
\end{eqnarray}
This completes the outline of the perturbative scheme as applied to our choice of metrics (\ref{metric}) and (\ref{gvmetric}). In the next section we will examine
the temporal aspect more closely.

\section{Explicit Form of the Temporal Function}

We employ the junction conditions derived by Maharaj et al. \cite{maharaj3} to determine the temporal evolution of our model.
\begin{equation}\label{bc}
\left({p}_r\right)_{\Sigma} = \left(qB\right)_{\Sigma} - \rho_s({\mathrm{r}},v)|_\Sigma
\end{equation}
It is important to point out that (\ref{bc}) holds only at the boundary of the star.
We require that the static pressure vanishes at the surface via the condition  $\left(p_{r0}\right)_{\Sigma} = 0$, so that the following equation
is obtained in $T(t)$, namely
\begin{equation} \label{bound}
\hspace{1cm} \alpha_{\Sigma}T - {\ddot T} =  2 \beta_{\Sigma} {\dot T} + \lambda_{\Sigma}
\end{equation}
where $\alpha$, $\beta$ and $\lambda$ are given by
\begin{eqnarray} \label{alp}
\alpha(r) &=& \frac{A_0^2}{B_0 b}\bigg[\left(\frac{B_0^{\prime}}{B_0} + \frac{1}{r} + \frac{A_0^{\prime}}{A_0}\right)\left(\frac{b}{B_0}\right)^{\prime} + \left(\frac{B_0^{\prime}}{B_0} + \frac{1}{r}\right)\left(\frac{a}{A_0}\right)^{\prime}  \nonumber \\
&& + B_0^2 \bigg(\left(\frac{3a}{2A_0} + \frac{b}{B_0}\right) p_{r0}  +  \left(\frac{3a}{2A_0} + \frac{2b}{B_0}\right)\rho_{s0} +    \nonumber \\
&& \left(\frac{a}{A_0} + \frac{b}{B_0}\right)\frac{3k}{2r B_0} \bigg) \bigg]
\end{eqnarray}
where $\rho_{s0}$ is the constant string density.
\begin{equation}
\hspace{1cm} \beta(r) = \frac{A_0^2}{2b}\left(\frac{b}{A_0B_0}\right)^{\prime}
\end{equation}
\begin{equation}
\lambda(r) = - \frac{A_0^2 B_0}{2\epsilon b} \left(\rho_s \right)
\end{equation}
to be evaluated at the boundary $r_\Sigma$. It should be noted that $p_{r0}$ vanishes at the boundary.  The diffusion equation (\ref{diff}) has
been extensively studied and  exact several solutions have been obtained \cite{glass,glass2}, \cite{maharaj3}, \cite{govin}, \cite{Ghosh1},
\cite{byron1} and \cite{NaiduNF1}.
One such solution of the diffusion equation (\ref{diff}) for which the string density is a function of the external radial coordinate is given by
\begin{equation} \label{ssss}
\rho_s({\mathrm{r}}) = \rho_0 + \frac{k}{{\mathrm{r}}}
\end{equation}
where $\rho_0$ and $k$ are constants. The string density profile in (\ref{ssss}) was utilised by Naidu et al. \cite{NaiduNF1} to study the effect of an anisotropic atmosphere on the temperature profiles during radiative collapse. The above choice of string profile generalizes earlier work by
Govender and Thirukannesh \cite{gov2} and Govender et al. \cite{BrasselB} in which the string density was constant ($k = 0$). The choice of a constant string density not only makes the problem mathematically tractable but also simplifies the underlying physics. The constant string distribution gives rise to pressure anisotropy in the exterior while any inhomogeneities are suppressed. Our choice (\ref{ssss}) allows for pressure anisotropy and inhomogeneities due to density fluctuations. At the boundary of the star the string density (\ref{ssss}) can be written as
\begin{equation} \label{zzzz}
\left(\rho_s\right)_\Sigma = \rho_0 + \frac{k}{rB}|_\Sigma
\end{equation}
where we have invoked the junction condition (\ref{4c}).

It is necessary to highlight the connection between $r$ and ${\mathrm{r}}$ at this point. The boundary of the collapsing star divides spacetime into two distinct regions ${\cal{M}}^{-}$ and ${\cal{M}}^{+}$, respectively. The coordinates in the interior spacetime $ {\cal{M}}^{-}$ are $(t, r, \theta,\phi)$ while the coordinates in $ {\cal{M}}^{+}$ are $(v, {\mathrm{r}}, \theta,\phi)$. The boundary $\Sigma$, is a time-like hypersurface described by the line element
\begin{equation}  \label{intrinsic}
ds^2_{\Sigma} = -d\tau^2 + {\cal{R}}^2(d\theta^2 + \sin^2{\theta}d\phi^2)
\end{equation}
endowed with coordinates $ {\xi}^i = (\tau, \theta, \phi)$ and ${\cal{R}} = {\cal{R}}(\tau) $.
Note that the time coordinate $\tau$ is defined only on the surface
$\Sigma$. The junction condition (\ref{4c}) is a consequence of requiring the smooth matching of the line elements (\ref{metric}) and (\ref{gvmetric})
across $ \Sigma$, ie.,
\begin{equation}   \label{int}
(ds^2_-)_{\Sigma} = (ds^2_+)_{\Sigma} = ds^2_{\Sigma}
\end{equation}
For ${\cal{M}}^{-}$  we obtain
\begin{eqnarray} \label{t1}
A(r_{\Sigma},t)\dot{t} &=& 1  \label{t1a} \\
r_{\Sigma}B(r_{\Sigma},t) &=& \cal{R}(\tau) \label{t1b}
\end{eqnarray}
where dots represent differentiation with respect to $ \tau$ while for ${\cal{M}}^{+}$ we have
\begin{eqnarray}
{{\mathrm{r}}}_{\Sigma}(v) &=& {\cal{R}}(\tau)  \label{t2a}  \\
\left( 1 - \displaystyle\frac{2m}{{\mathrm{r}}} +  2\displaystyle\frac{d{{\mathrm{r}}}}{dv}
\right)_{\Sigma} &=& \left( \frac{1}{{\dot{v}}^2}\right)_{\Sigma}
\label{t2b}
\end{eqnarray} We observe that (\ref{t1b}) and (\ref{t2a}) relate $r$ and ${\mathrm{r}}$.

We complete the expression for the temporal function $ T(t)$ by solving (\ref{bound}).
This gives
\begin{equation} \label{temporal}
\hspace{1cm} T(t) =  \frac{\lambda_{\Sigma}}{\alpha_{\Sigma}} - \exp\left[\left(-\beta_{\Sigma}+\sqrt{\alpha_{\Sigma}+\beta^2_{\Sigma}}\right)t\right]
\end{equation}
which, together with (\ref{alp}) and (\ref{ssss}) and $\alpha_{\Sigma} > 0$ as well as $\beta_{\Sigma} < 0$ describes a system in static equilibrium that starts to collapse at $t = -\infty$ and continues to collapse as $t$ increases.

\section{Dynamical Model}
\label{model}
In order to investigate the properties of the extended form of the temporal function, we make use of the simple Schwarzschild interior metric in
isotropic coordinates \cite{bon,ray} given by
\begin{eqnarray} \label{9}
A_0(r) &=&  c_1 - \frac{1}{2}\frac{\left(1-r^2\right)}{(1+r^2)} \label{9a} \\
B_0(r) &=& \frac{2R}{1+r^2} \label{9b}
\end{eqnarray}
where $c_1$ and $R$ are constants. Then (\ref{6a}) and (\ref{6b}) can be written as
\begin{eqnarray} \label{6'}
\mu_0 &=& \frac{3}{R^2} \label{6a'} \\
p_{r0} &=& -\frac{2c_1(1+r^2) - 3(1-r^2)}{R^2\left[2c_1(1+r^2) - (1-r^2)\right]} \label{6b'}
\end{eqnarray}
The constant $R$ is easily determined from (\ref{6a'}), given the initial static energy density, and parameter $c_1$ is obtained from (\ref{6b'}) by
evaluation at the boundary, giving
\begin{equation} \label{6c'}
c_1 = \frac{3(1 - r_\Sigma^2)}{2(1 + r_\Sigma^2)}
\end{equation}
Restrictions on $r_\Sigma$ as given by Santos \cite{bon} are noted, namely
\begin{equation}
\frac{2m_0}{r_\Sigma} = \frac{4 r_\Sigma^2}{(1+r_\Sigma^2)^2} < \frac{3}{4}
\end{equation}
and
\begin{equation}
0 \leq r < r_\Sigma < \frac{1}{\sqrt{3}}.
\end{equation}
We also note that in the case $p_r = p_t$, the anisotropy parameter $\Delta$ vanishes.

\section{Radiating Collapse}
\label{radiate}
We note that (\ref{7a})-(\ref{7d}) contain two unspecified quantities, namely $a(r)$ and $b(r)$, which modulate the temporal part of the gravitational potentials. Thus it is important that these are determined carefully in order to obtain a physically meaningful dynamical model. Following Chan et al. \cite{chan2} we adopt the following form for $b(r)$,
\begin{equation}
\hspace{1cm} b(r) = \left(1+\xi f(r)\right)A_0B_0 \label{bee}
\end{equation}
This choice for $b(r)$ has been widely used to investigate stability of radiating stars undergoing dissipative collapse in the form of a radial
heat flux \cite{Kich}, \cite{herr2} and \cite{chan2}. Furthermore, we follow \cite{narenee} and choose the following form $f(r) = r^2$.
Using (\ref{bee}) in (\ref{pertiso}) above, we obtain an explicit form for $a(r)$ as
\begin{eqnarray}
a(r) &=& \left(\frac{2c_1 +1}{2}  - \frac{1}{1+r^2}\right)\times \bigg[c_2 - \frac{1-\xi}{1+r^2} - \frac{\xi}{2}(2c_1 + 1)(1+r^2) \bigg]  \nonumber \\
&&  + \frac{(2c_1+1)(1 + 2c_1(1-\xi) - 11\xi) - 8c_3 R^2}{2(2c_1+1)(1+r^2)} +  \nonumber \\
&& 2\xi \left(2c_1+1\right)\log(1+r^2)
\end{eqnarray}
where $c_2$ and $c_3$ are constants of integration. These may be set by considering the work of Govender et al. \cite{gov1} with the simple case of
the relationship between $a(r)$ and $b(r)$ being employed, namely
\begin{equation}
\frac{a(r)}{A_0(r)} = \frac{b(r)}{B_0(r)}
\end{equation}
At this stage, we point out that the radial and temporal evolution of our model is fully determined. We use numerical data given in Table 1. for
performing graphical analyses of stability and horizon formation comparisons which follow.
\begin{table}[ht]\caption{Values of constants used in model}
\centering
\begin{tabular}{c c c}
\hline\hline
Name & Value & Reference\\ [0.5ex]
\hline
$\mu_0$ & $2.36321 \times 10^8 g/cm^3$ & \cite{bon} \\
$r_\Sigma$ & $2.159 \times 10^8 cm$ & \cite{bon} \\
R & $26.08 \times 10^8 cm$ & (\ref{6a'}) \\
$c_1$ & 1.49984 & (\ref{6c'}) \\
$\xi$ & -0.2 & \cite{chan2,narenee} \\
[1ex]
\hline
\end{tabular}\label{table:constants}
\end{table}

\subsection{Luminosity}
The luminosity for an observer at rest at infinity is given by
\begin{equation} \label{lumin1}
L_\infty (v)= - \left(\frac{dm(v)}{dv}\right)_\Sigma
\end{equation}
where $v$ is the retarded time. The luminosity can then be written as
\begin{equation} \label{lumin2}
- \frac{dm(v)}{dv}= - \frac{dm}{dt} \frac{dt}{d\tau} \frac{d\tau}{dv}
\end{equation}
where
\begin{equation}
\frac{dt}{d\tau} = \frac{1}{A} \hspace{0.6cm} \mathrm{and} \hspace{0.6cm} \frac{d\tau}{dv} = \frac{1}{\dot{v}} =  \frac{r \dot{B}}{A} + \frac{B + r B^\prime}{B}
\end{equation}
and $\tau$ is the proper time defined on $\Sigma$ \cite{pin} with $A$ and $B$ given by (\ref{5a}) and (\ref{5b}). For our model, this is calculated
to be
\begin{equation}
L_\infty (t) = - \frac{\epsilon g_{r_\Sigma} \dot{T}(t)}{A_0 + \epsilon a T(t)} \left( 1 + \frac{r  B_0^{\prime}}{B_0 + \epsilon b T(t)} \right) |_{r=r_\Sigma}  \label{lumin}
\end{equation}
where
\begin{equation}
g_{r_\Sigma} = - r^2 b^{\prime} - \frac{r^3{{B_0^{\prime}}^2}}{2B_0} \left(2\frac{b^{\prime}}{B_0^{\prime}}-\frac{b}{B_0}\right) |_{r=r_\Sigma}
\end{equation}

\subsection{Horizon Formation}
From (\ref{lumin1}) and (\ref{lumin2}) we note that the luminosity vanishes when
\begin{equation}
\frac{1}{\dot{v}} = 0
\end{equation}
which determines the time of formation of the horizon as the collapse process proceeds from $-\infty < t \leq t_H$ which corresponds
to $-\infty < v < \infty$. For our model we have the following instances where the luminosity vanishes. By examining (\ref{lumin}) we must
have $\dot{T} = 0$ or $g_{r_\Sigma} = 0$.
\subsubsection{Case 1: $\dot{T} = 0$}
In the case $\dot{T} = 0$, we have from the derivative of (\ref{temporal})
\begin{equation}
- \eta e^{\eta t} = 0
\end{equation} \label{eta}
where
\begin{equation}
\eta = - \beta_{\Sigma} + \sqrt{\alpha_{\Sigma} + \beta_{\Sigma} ^2}
\end{equation}
Thus $\dot{T} = 0$ implies $\eta = 0$ which forces $\alpha_{\Sigma} = 0$. From the expression for $T(t)$ (\ref{temporal}), this is only possible
if $\lambda_{\Sigma} = 0$ (ie. vanishing of the string density). We observe that removing the string density gives a pure radiation solution and
the horizon is able to form. However, the inclusion of strings inhibits the formation of the horizon.
\subsubsection{Case 2: $g_{r_\Sigma} = 0$}
The second case $g_{r_\Sigma} = 0$ gives
\begin{equation} \label{grrzero}
\frac{2 r^3 R h(r)}{(1 + r^2)^4} = 0
\end{equation}
using $B_0(r)$ from (\ref{9b}) and $b(r)$ from (\ref{bee}) where we have used
\begin{eqnarray}
h(r) &=& (r^2 - 1)(3 + \xi (r^4 + 3 r^2 - 1)) + 2 c_1 (1 + r^2 - \xi + 2 r^4 \xi + r^6 \xi)
\end{eqnarray}
Examining (\ref{grrzero}) we see that $g_{r_\Sigma} = 0$ when either $R = 0$ or $h(r) = 0$. For our model, we have chosen an initial
density $\mu_0$ as given in Table 1. following numerical work by Santos \cite{bon}. By (\ref{6a'}) this means $R \neq 0$ hence we consider the second possibility of $h(r) = 0$.
Given $R$, this places a restriction on $c_1$ thus creating a relationship between constants $c_1$ and $R$.

\section{Stability Analysis}
\label{stability}
In order to provide insight into the stability of the star, we begin with the second law of thermodynamics, and follow the approach
of \cite{herr2}. This leads to the adiabatic parameter $\Gamma$ which is the ratio of specific heats at constant pressure and constant
volume, and taken to be constant throughout the distribution (or at least in the region being studied). In literature referring to this
ratio in \cite{herr2}, it was considered an indicator of stability, and called the stability factor \cite{narenee}. From the expressions
for $\Gamma$ given below \cite{chan2}, it is clear that pressure anisotropy and the presence of radiation within the stellar core affect the stability
factor. For example, if the sign of the anisotropy parameter $\Delta = (p_{t0} - p_{r0})$ changes, the stellar core becomes unstable. We
observe this is in the Newtonian limit,
\begin{eqnarray}
\Gamma &<& \frac{4}{3}+\left[\frac{1}{3|p_{r0}^{\prime}|}\left(4 \frac{(p_{t0} - p_{r0})}{r} + 2 \alpha_{\Sigma}\xi |f^{\prime}| \right) \right]_{max} \label{11a}
\end{eqnarray}
In agreement with classical fluid dynamics, the fluid sphere becomes more unstable (increasing the unstable range of $\Gamma$) as a result of the
Newtonian contribution due to dissipation.
Relativistic contributions from the energy density lead to a stability factor different from its Newtonian counterpart.
\begin{eqnarray}
\Gamma &<& \frac{4}{3}+\bigg[ \frac{1}{3|p_{r0}^{\prime}|}\bigg( 4\frac{(p_{t0} - p_{r0})}{r} + 2\alpha_{\Sigma}\xi |f^{\prime}| + \mu_0\bigg( p_{r0} r  - \xi|f^{\prime}| \bigg) \bigg)\bigg]_{max} \label{11b}
\end{eqnarray}
Equation (\ref{11b}) shows that the unstable range of $\Gamma$ is increased by the Newtonian term due to dissipation, as in the Newtonian limit.
Furthermore, the unstable range of $\Gamma$ is increased by the relativistic correction due to the static background fluid
configuration; however the relativistic correction due to dissipation decreases the unstable range of $\Gamma$. Bonnor et. al. \cite{bon} state
that dissipation, by diminishing that total mass entrapped inside the fluid sphere, renders the systems less unstable.

In order to investigate the stability of our model in both the Newtonian and post-Newtonian limits, we graphed $\Gamma$ for the case of pure
radiation (absence of string density), radiation plus constant density ($k = 0$) and the radiation and inhomogeneous string
density ($\rho_s \neq 0 , k \neq 0$). Since our static model is described by the interior Schwarzschild solution, the anisotropy
parameter $\Delta = p_{t0} - p_{r0}$ vanishes in (\ref{11a}) and (\ref{11b}). The modified effects due to pure string density and
inhomogeneity are encoded in $\alpha_\Sigma$. We have also graphed the luminosity as a function of time for both the pure and generalized
Vaidya exterior. It is important to note that the graphs are plotted using geometrized units, where $G$ and $c$ are taken to be unity \cite{BrasselB}.
\begin{figure}
	\centering
	\includegraphics[scale=0.65]{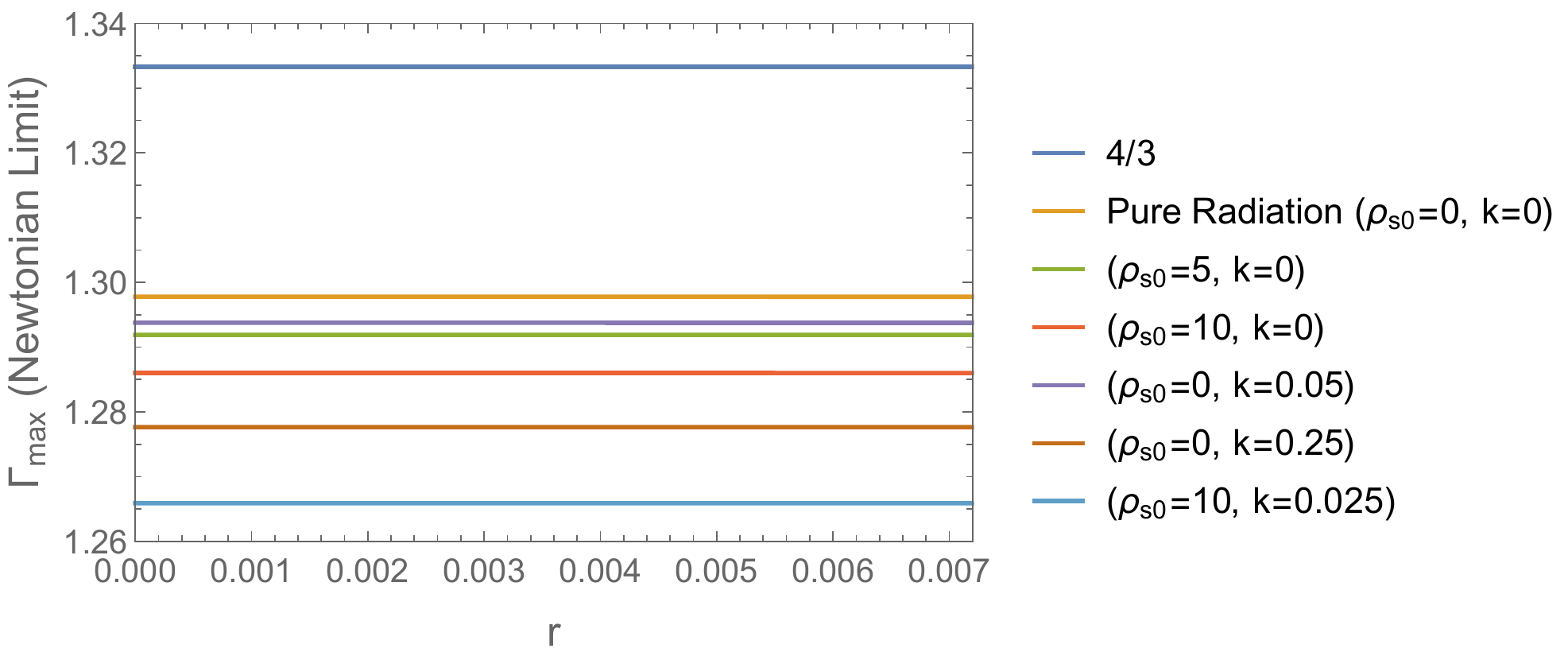}
	\caption{Adiabatic parameter in the Newtonian limit for various string densities, $\rho_s = \rho_s(\rho_{s0}, k)$ from the centre $r = 0$ to the boundary $r = r_\Sigma$, $r$ in units of seconds.} \label{fig1}
\end{figure}
\begin{figure}
	\centering
	\includegraphics[scale=0.65]{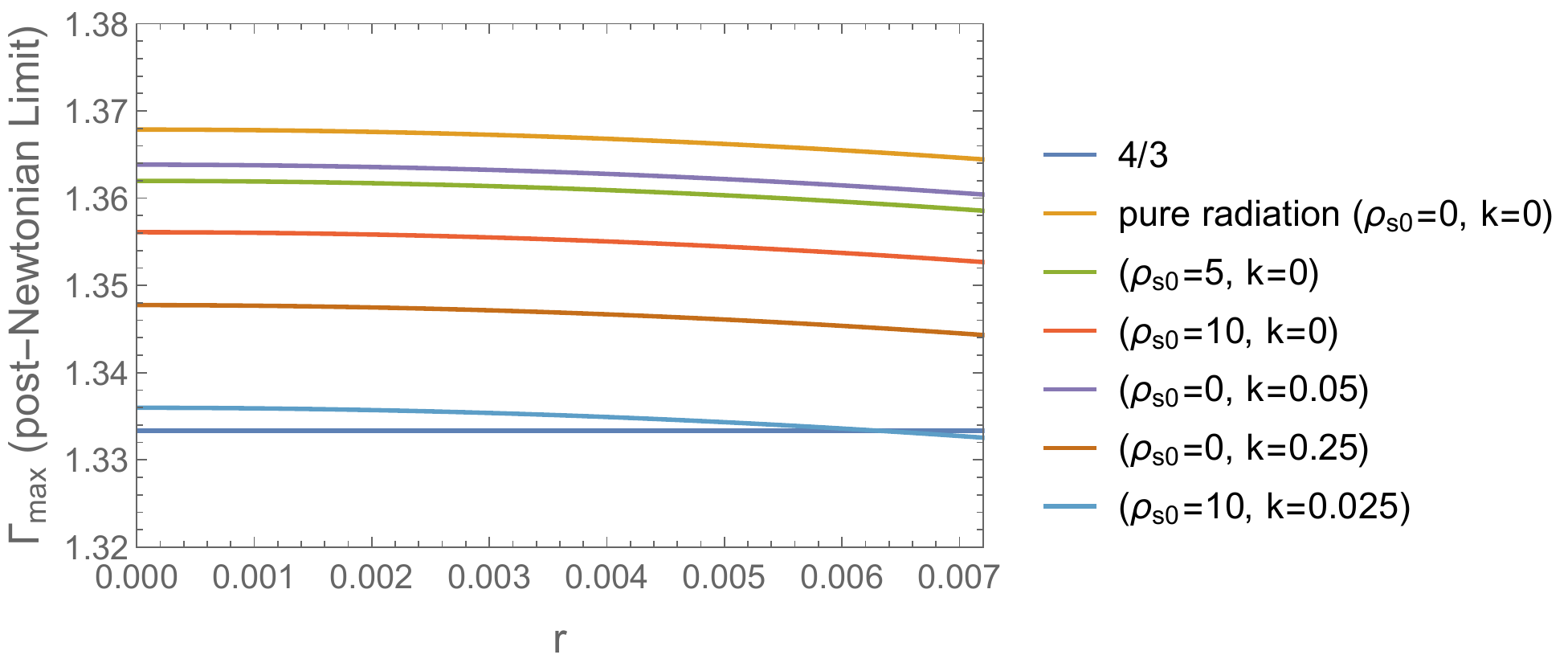}
	\caption{Adiabatic parameter in the post-Newtonian limit for various string densities, $\rho_s = \rho_s(\rho_{s0}, k)$ from the centre $r = 0$ to the boundary $r = r_\Sigma$, $r$ in units of seconds.} \label{fig2}
\end{figure}
\begin{figure}
	\centering
	\includegraphics[scale=0.65]{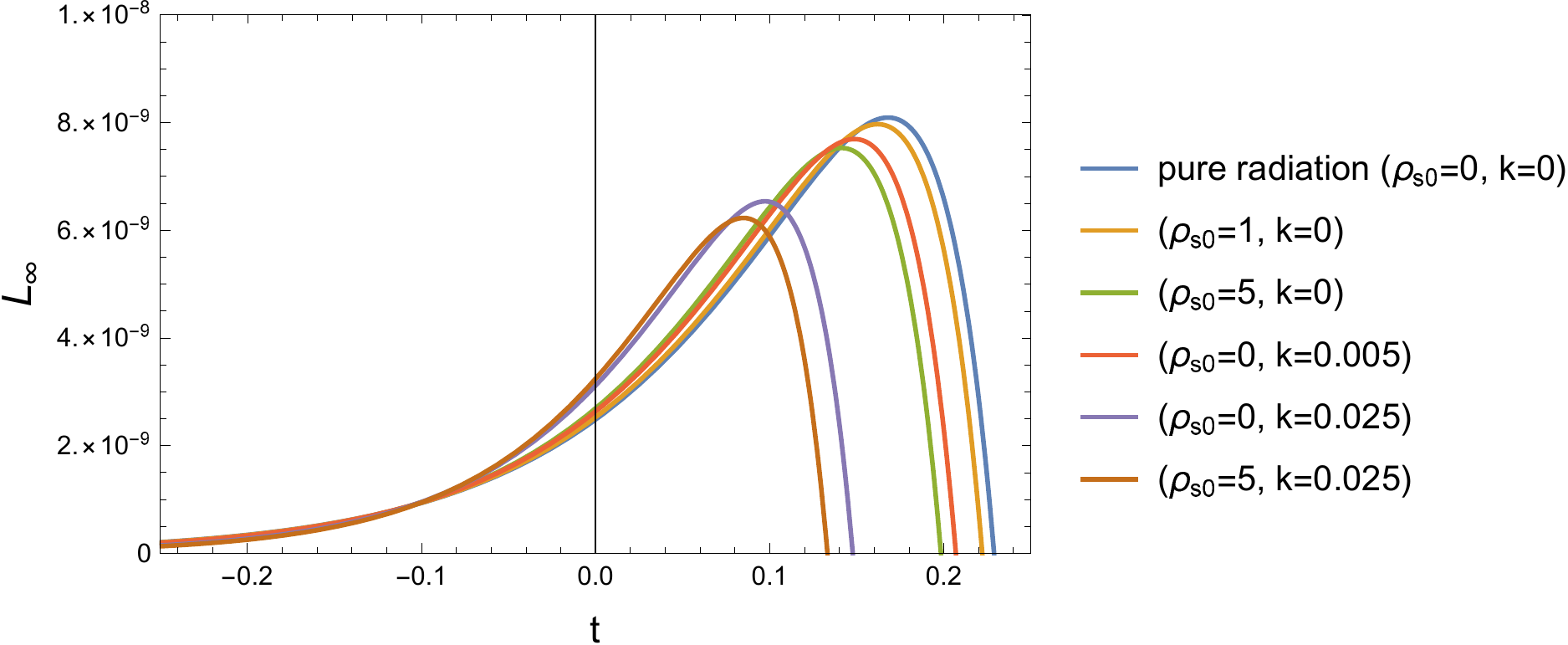}
	\caption{Luminosity at infinity (geometrized) versus time (seconds) for various string densities, $\rho_s = \rho_s(\rho_{s0}, k)$}. \label{fg3}
\end{figure}

\section{Physical Analysis and Conclusion}
\label{analyse}
Figure 1 shows the stability factor when the star is close to hydrostatic equilibrium in the Newtonian
limit. We observe that the different matter configurations exhibit instability with $\Gamma < \frac{4}{3}$ which
signifies the onset of collapse. The inclusion of the string field drives the stellar fluid towards
instability with this effect being enhanced by inhomogeneity ($k > 0$).
Figure 2 displays $\Gamma$ for the post-Newtonian regime. It is clear that the collapse process drives the
fluid towards stability. The presence of the strings and their associated anisotropy and inhomogeneity make the fluid more stable at late times. This could be due to trapping of heat within the stellar core due to an inhomogeneous atmosphere, thus resulting in higher core temperatures. An increase in the core temperature results in an increase in outward pressure thus hindering gravitational collapse.
In Figure 3 we note that inclusion of the string density field promotes an earlier time of horizon formation (luminosity vanishes),
with this effect being particularly sensitive to string inhomogeneity. The effect of the string density on
the time of formation of the horizon was also observed by Govender \cite{megan}. They reasoned that the presence of the anisotropic strings in the exterior lowered the rate of heat dissipation to the exterior thus leading to a lower heat production rate within the core. This results in a lower outward radial pressure thus allowing gravity to dominate within the stellar interior. This results in a higher collapse rate and eventually to the horizon forming earlier. Luminosities for collapsing neutron star models were studied by de Oliveira et al. \cite{de1} in which they considered profiles in the $\gamma$-ray, X-ray and visible bandwidths for core masses of $2M_{\odot}$ and $3M_{\odot}$ and a radius of 10 km. The luminosity profiles for all three bandwidths are similar to the profiles depicted in Figure 3. They noted that the radiation pulses do not occur simultaneously for an observer placed at infinity from the collapsing body. Furthermore, they show that nearly all the energy is emitted in the form of $\gamma$-rays. It is well known that the luminosity profile depends on the increasing gravitational redshift as the star collapses and the increase in the effective temperature. Our study provides a possible mechanism to explain the temperature changes within the core which manifests as the luminosity profiles displayed in Figure 3.

It is important to note that while the dynamical (in)stability of radiating spheres has been extensively studied in the Newtonian and post-Newtonian approximations, our investigation is the first attempt at considering the dynamics of the collapse process with a generalised Vaidya exterior. The generalised Vaidya atmosphere alters the temporal evolution of the model which impacts on the stability and time of formation of the horizon. While our study has considered an imperfect fluid interior with heat dissipation and pressure anisotropy it would be interesting to include shear viscosity within the framework of extended irreversible thermodynamics.

\section{Acknowledgments}
NFN wishes to acknowledge funding from the National Research Foundation (Grant number: 116629) as well as the DSI-NRF Centre of Excellence in Mathematical and Statistical Sciences (CoE-MaSS). RB and MG acknowledge financial support from the office of the DVC: Research and Innovation at the Durban University of Technology. MG is indebted to Dr A Barnes for providing the excellent facilities at Glenwood Boys High School where part of this work was completed.

\label{lastpage}


\begin{thebibliography}{}
\bibitem{oppsnyd} Oppenheimer, J.R., Snyder, H.: \textsf{On continued gravitational contraction}. Phys. Rev. D. {\bf 56}, 455 (1939)
\bibitem{vaidya} Vaidya, P.C.: \textsf{The gravitational field of a radiating star}. Proc. Indian Acad. Sc. A. {\bf 33}, 264 (1951)
\bibitem{santos} Santos, N.O.: \textsf{Non-adiabatic radiating collapse}. Mon. Not. R. Astron. Soc. {\bf 216}, 403 (1985)
\bibitem{mken} Mkenyeleye, M.D., Goswami, R., Maharaj, S.D.: \textsf{Gravitational collapse of generalised Vaidya spacetime}. Phys. Rev. D. {\bf 92} 024041 (2015)
\bibitem{sharma1} Sharma, R., Tikekar, R.: \textsf{Non-adiabatic radiative collapse of a relativistic star under different initial conditions}. Pramana-J. Phys. {\bf 79}, 501 (2012)
\bibitem{sharma2} Sharma, R., Tikekar R.: \textsf{Space-time inhomogeneity, anisotropy and gravitational collapse}. Gen. Relativ. Gravit. {\bf 44}, 2503-2520 (2012)
\bibitem{sharma3} Sharma, R., Das, S., Rahaman, F., Shit, G.C.: \textsf{Gravitational collapse of a circularly symmetric star in an anti-de Sitter spacetime}. Astrophys. Space Sci. {\bf 259}, 40 (2015)
\bibitem{gov3} Govender, M., Bogadi, R., Sharma, R., Das, S.: \textsf{Gravitational collapse in spatially isotropic coordinates}. Astrophys. Space Sci. {\bf 361}, 33 (2016)
\bibitem{kevin} Govender, M., Reddy, K.P., Maharaj, S.D.: \textsf{The role of shear in dissipative gravitational collapse}. Int. J. Mod. Phys. D. {\bf 23}, 1450013 (2014)
\bibitem{chandra} Chandrasekhar, S.: \textsf{The dynamical instability of gaseous masses approaching the Schwarzschild limit in general relativity}. Astrophys. J. {\bf 140}, 417 (1964)
\bibitem{herr2} Herrera, L., Le Denmat, G., Santos, N.O.: \textsf{Dynamical instability for non-adiabatic spherical collapse}. Mon. Not. R. Astron. Soc. {\bf 237}, 257 (1989)
\bibitem{chan2} Chan, R., Herrera, L., Santos, N.O.: \textsf{Dynamical Instability for Radiating Anisotropic Collapse}. Mon. Not. R. Astron. Soc. {\bf 265}, 533 (1993)
\bibitem{herr3} Herrera, L., Le Denmat, G., Santos, N.O.: \textsf{Dynamical instability and the expansion-free condition}. Gen. Relativ. Gravit. {\bf 44}, 1143 (2012)
\bibitem{bon} Bonnor, W.B., de Oliveira, A.K.G., Santos, N.O.: \textsf{Radiating spherical collapse}. Phys. Rep. {\bf 181}, 269 (1989)
\bibitem{glass} Glass, E.N., Krisch, J.P.: \textsf{Radiation and string atmosphere for relativistic stars}. Phys. Rev. D. {\bf 57}, 5945 (1998)
\bibitem{glass2} Glass, E.N., Krisch, J.P.: \textsf{Two-fluid atmosphere for relativistic stars}. Class. Quantum Grav. {\bf 16}, 1175 (1999)
\bibitem{maharaj2} Maharaj, S.D., Govender, M.: \textsf{Radiating Collapse with Vanishing Weyl stresses}. Int. J. Mod. Phys. D. {\bf 14}, 667-676 (2005)
\bibitem{maharaj3} Maharaj, S.D., Govender, G., Govender, M.: \textsf{Radiating stars with generalised Vaidya atmospheres}. Gen. Relativ. Gravit. {\bf 44}, 1089-1099 (2012)
\bibitem{HusV} Husain, V.: \textsf{Exact solutions for null fluid collapse}. Phys. Rev. D. {\bf 53}, 1759 (1996)
\bibitem{WangA} Wang, A., Wu, Y.: \textsf{Generalized Vaidya Solutions}. Gen. Relativ. Gravit. {\bf 31}, 107 (1999)
\bibitem{govin} Govinder, K.S., Govender, M.: \textsf{Gravitational collapse of null radiation and a string fluid}. Phys. Rev. D. {\bf 68}, 024034 (2003)
\bibitem{oliv2} de Oliveira, A.K.G., Santos, N.O., Kolassis, C.A.: \textsf{Collapse of a radiating star}. Mon. Not. R. Astron. Soc. {\bf 216}, 1001 (1985)
\bibitem{gov2} Govender, M., Thirukkanesh, S.: \textsf{Anisotropic static spheres with linear equation of state in isotropic coordinates}. Astrophys. Space Sci. {\bf 358}, 16 (2015)
\bibitem{kevin1} Reddy, K.P., Govender, M., Maharaj, S.D.: \textsf{Impact of anisotropic stresses during dissipative gravitational collapse}. Gen. Relativ. Gravit. {\bf 47}, 35 (2015)
\bibitem{Ghosh1} Ghosh, S.G., Deshkar, D.W.: \textsf{Exact Nonspherical Radiating Collapse}. Int. J. Mod. Phys. A. {\bf 22}, 2945 (2007)
\bibitem{byron1} Brassel, P.B., Goswami, R., Maharaj, S.D.: \textsf{Collapsing radiating stars with various equations of state}. Phys. Rev. D. {\bf 95}, 124051 (2017)
\bibitem{NaiduNF1} Naidu, N.F., Govender, M., Thirukkanesh, S., Maharaj, S.D.: \textsf{Radiating fluid sphere immersed in an anisotropic atmosphere}. Gen. Relat. Gravit. {\bf 49}, 95 (2017)
\bibitem{BrasselB} Govender, G., Brassel, P.B., Maharaj, S.D.: \textsf{The effect of a two-fluid atmosphere on relativistic stars}. Eur. Phys. J. C. {\bf 75}, 324 (2015)
\bibitem{ray} Raychaudhuri, A.K., Maiti, S.R.: \textsf{Conformal flatness and the Schwarzschild interior solution}. J. Math. Phys. {\bf 20}, 245 (1979)
\bibitem{Kich} Chan, R., Kichenassamy, S., Le Denmat, G., Santos, N.O.: \textsf{Heat flow and dynamical instability in spherical collapse}. Mon. Not. R. Astron. Soc. {\bf 239}, 91 (1989)
\bibitem{narenee} Govender, M., Mewalal, N., Hansraj, S.: \textsf{The role of an equation of state in the dynamical (in)stability of a radiating star}. Eur. Phys. J. C. {\bf 79}, 24 (2019)
\bibitem{gov1} Govender, M., Govinder, K.S., Maharaj, S.D., Sharma, R., Mukherjee, S., Dey, T.K.: \textsf{Radiating Spherical Collapse with Heat Flow}. Int. J. Mod. Phys. D. {\bf 12}, 667 (2003)
\bibitem{pin} Pinheiro, G., Chan, R.: \textsf{Radiating gravitational collapse with shear viscosity revisited}. Gen. Relativ. Gravit. {\bf 40}, 2149 (2008)
\bibitem{megan} Govender, M.: \textsf{Nonadiabatic Spherical Collapse with a Two-Fluid Atmosphere}. Int. J. Mod. Phys. D. {\bf 22}, 1350049 (2013)
\bibitem{de1} de Oliveira, A.K.G., de F. Pacheco, J.A., Santos, N.O.: \textsf{More about collapse of a radiating star}. MNRAS. {\bf 220}, 405 (1986)
\end{thebibliography}
\end{document}